\documentclass{article}
\usepackage{spconf,amsmath,hyperref}
\usepackage[pdftex]{graphicx}
\usepackage{amssymb}
\usepackage{booktabs}
\usepackage{subcaption}
\usepackage{multirow}
\usepackage{xcolor}
\usepackage{siunitx}
\usepackage{cite}
\newcommand{\hoa}[1]{\textcolor{gray}{#1}}
\usepackage{array, amsmath, amssymb, amsfonts, bm, mleftright, xargs}

\newcommand{\Hr}{\mathsf{H}}

\newcommand{\mbA}{\mathbf{A}}

\newcommand{\mba}{\mathbf{a}}

\newcommand{\mbx}{\mathbf{x}}

\title{
SIRUP: A Diffusion-Based Virtual Upmixer of Steering Vectors\\
for Highly-Directive Spatialization with First-Order Ambisonics
\vspace{-1mm}
}
\name{Emilio Picard\textsuperscript{1,2} \ \
Diego Di Carlo\textsuperscript{1} \ \
Aditya Arie~Nugraha\textsuperscript{1} \ \
Mathieu Fontaine\textsuperscript{3,1} \ \
Kazuyoshi Yoshii\textsuperscript{4,1}
\vspace{-1mm}
\thanks{This work was supported by JST FOREST no. JPMJFR2270, 
JSPS KAKENHI nos. JP23K16912, JP23K16913, 24H00742, 24H00748, and 25H01142, and ANR Project SAROUMANE (ANR-22-CE23-0011). 
All the code used to produce the results of this paper 
is available at \texttt{\url{https://github.com/emilio-pcrd/sirup}} upon acceptance.}}
\address{
\textsuperscript{1}Center for Advanced Intelligence Project, RIKEN, Japan \\
\textsuperscript{2}Sorbonne University, France \\
\textsuperscript{3}LTCI, T\'{e}l\'{e}com Paris, Institut Polytechnique de Paris, France \\
\textsuperscript{4}Graduate School of Engineering, Kyoto University, Japan
\vspace{-1mm}
}
\begin{document}
\fussy
\setlength{\abovedisplayskip}{5pt}
\setlength{\belowdisplayskip}{5pt}
\setlength{\abovedisplayshortskip}{5pt}
\setlength{\belowdisplayshortskip}{5pt}
\allowdisplaybreaks[4]

\maketitle
\begin{abstract}
\vspace{-.6mm}
This paper presents virtual upmixing of steering vectors 
 captured by a fewer-channel spherical microphone array. 
This challenge has conventionally been addressed 
 by recovering the directions and signals of sound sources 
 from first-order ambisonics (FOA) data, 
 and then rendering the higher-order ambisonics (HOA) data 
 using a physics-based acoustic simulator. 
This approach, however, struggles to handle 
 the mutual dependency between
 the spatial directivity of source estimation
 and the spatial resolution of FOA ambisonics data.
Our method, named SIRUP, 
 employs a latent diffusion model architecture.
Specifically, a variational autoencoder (VAE) 
 is used to learn a compact encoding of the HOA data in a latent space
 and a diffusion model is then trained to generate the HOA embeddings, 
 conditioned by the FOA data. 
Experimental results showed 
 that SIRUP achieved a significant improvement compared to FOA systems 
 for steering vector upmixing, source localization, and speech denoising.
\end{abstract}
\vspace{-.3mm}
\vspace{-.7mm}
\section{Introduction}
\label{sec:intro}
\vspace{-1mm}

Sound source localization (SSL) 
 and speech enhancement (SE) or denoising 
 remain fundamental tasks in machine listening applications, 
 further used in augmented reality (AR) scenes \cite{9770284}, 
 as well as robotics \cite{10551479}, 
 radar \cite{doi:10.1049/ip-f-1.1980.0041}, 
 and autonomous driving systems \cite{electronics11050766}.
Beamforming and SSL methods are still widely characterized by acoustic and signal processing ~\cite{DiBiase2001, book_vincent, 1143830}, even though recent work tried to incorporate deep neural networks (DNNs) to address such tasks ~\cite{adavanne2021differentiabletrackingbasedtrainingdeep, 8567942}.
These applications call for more accurate spatial information that could be translated as more selective beam patterns, enabling better localization and spatial filtering.

High-order-ambisonic (HOA) microphone arrays \cite{ambisonics1} are capable of high spatial resolutions, enabling effective sound analysis and synthesis~\cite{bertet2012doahoacomparaison, avni_spatial_2013}.
The spatial resolution of these systems does
relate to the number of ambisonic orders, which are related to the number of microphones. These limitations can be alleviated by adopting HOA setup, albeit at the expense of expensive hardware; consequently, common spatial recording systems are typically constrained to the first order (FOA), employing four microphones.

\begin{figure}
    \centering
    \includegraphics[width=.85\linewidth, trim=0cm .55cm 0cm .5cm, clip]{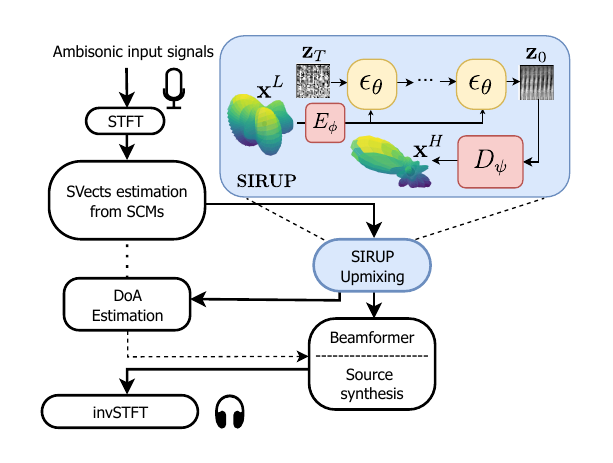}
    \caption{The SIRUP upmixer for downstream tasks.}
    \label{fig:pipeline}
    \vspace{-1em}
\end{figure}

A common remedy is parametric \emph{upmixing},
 which estimates scene parameters, 
 e.g., directions of arrival (DOA) via SSL 
 and source signals via separation or SE, 
 and then render virtual HOA channels. 
For example, major methods such as
 DirAC~\cite{Fuchs_2025} and COMPASS~\cite{compass}
 encode directional and ambient components 
 before ad-hoc HOA spatial rendering~\cite{jarrett2012rigid}.
Their performance is tightly coupled to the accuracy of steering vectors (SVs), 
 the spatial signatures driving classical SSL (e.g., SRP-PHAT) 
 and SE (e.g., delay-and-sum beamforming)~\cite{gannot2017consolidated}. 
This cascaded analysis–rendering pipeline is brittle: 
 low-resolution noisy FOA SVs degrade the parameter estimates 
 and propagate errors to the HOA rendering.

We instead propose to directly upmix the \emph{ambisonic SVs}.
SVs compactly encode the spatial characteristics of sources, 
 encompassing both the direct paths and early reflections. 
By super-resolving FOA SVs, 
 one can perform SSL in a highly-directive HOA space, 
 allowing the DOA estimates 
 to produce sharper spatial filters. 
The upmixed SVs can also be utilized
 to render the source images at a higher spatial resolution.

Specifically, a SteerIng vectoR UPmixer (\texttt{SIRUP}) learns a latent space of HOA SVs with a VAE
 and uses a conditional diffusion model 
 to generate HOA embeddings from FOA inputs estimated 
 under an isotropic ambient prior.
Once DOAs are inferred from the reconstructed HOA SVs, 
 source signals are recovered via beamforming
 and subsequently rendered using 
 FOA-consistent or either algebraic free-field SVs, 
 as depicted in Fig.~\ref{fig:pipeline}.
 
\vspace{-2mm}
\section{Background}
\vspace{-1mm}

This section introduces SVs and their applications. 
We work in the short-time Fourier transform (STFT) domain
 whose frequency bins and time frames are indexed
 with \(f\in\{1,\ldots,F\}\) and \(t\in\{1,\ldots,T\}\), 
 where \(F\) and \(T\) are the numbers of frequency bins and frames, respectively.

\vspace{-2mm}
\subsection{Steering vectors}
\label{sec:svect}
\vspace{-1mm}

Suppose we have an $M$-channel STFT observation $\mathbf{x}_{ft}\in\mathbb{C}^M$ of a static, far-field source 
\cite{book_vincent},
which is given by
\begin{equation}
\mathbf{x}_{ft}=\mathbf{a}_{f}\,s_{ft}+\mathbf{n}_{ft},
\end{equation}
where $\mathbf{a}_{f}\in\mathbb{C}^M$ is the acoustic transfer function, $s_{ft}\in\mathbb{C}$ the source STFT, and $\mathbf{n}_{ft}\in\mathbb{C}^M$ diffuse isotropic noise.

For a given array geometry and look direction $\theta$, the \textit{algebraic} SVs that models the direct path of $\mba_f$ is~\cite{book_vincent}
\begin{equation}
\tilde{\mathbf{a}}_{f}(\theta)
=\big[\,e^{-j2\pi f\,\tau_{1}(\theta)},\ldots,e^{-j2\pi f\,\tau_{M}(\theta)}\,\big]^{\top},
\label{eq:algebraic_svects}
\end{equation}
with relative delays $\tau_{m}(\theta)=\mathbf{u}_\theta^{\top}(\mathbf{r}_{m}-\bar{\mathbf{r}}) / c$,
where $\mathbf{r}_{m}$ is the $m$-th microphone position, $\bar{\mathbf{r}}$ the array center, $\mathbf{u}_\theta$ the unit vector pointing to $\theta$, and $c$ the speed of sound.




The dominant spatial mode of the spatial covariance matrix (SCM) 
 provides a good estimate of the SVs~\cite{markovich2009multichannel}:
\begin{equation}
\hat{\boldsymbol{\Sigma}}_f=\frac{1}{T}\sum_{t=1}^{T}\mathbf{x}_{ft}\mathbf{x}_{ft}^{\mathrm H}, 
\qquad 
\hat{\mathbf{a}}_f=\operatorname{eig}_1\!\left(\hat{\boldsymbol{\Sigma}}_f\right),
\label{eq:measured_svect}
\end{equation}
where $\operatorname{eig}_1(\cdot)$ denotes the principal eigenvector. These \emph{measured} SVs capture the direct path plus early reflections, and thus are richer than the algebraic model in Eq.~\eqref{eq:algebraic_svects}. 

\vspace{-2mm}
\subsection{Applications}
\vspace{-1mm}

For SSL, we evaluate 
 the steered-response-power (SRP) map
 on a discrete grid of DOAs and pick one $\hat\theta$
 \cite{grinstein2024steered} as follows:
\begin{equation}
\mathcal{S}(\theta)=
\sum_f 
    \left|\,
        \frac{\tilde{\mathbf{a}}_f(\theta)^\Hr}{\|\tilde{\mathbf{a}}_f(\theta)\|_2}
        \,
        \frac{\hat{\mathbf{a}}_f}{\|\hat{\mathbf{a}}_f\|_2}\,
    \right|^2,
\,
\hat{\theta}=\arg\max_\theta \mathcal{S}(\theta).
\label{eq:ssl}
\end{equation}
This corresponds to cross-correlation 
 between the estimated SV and algebraic SVs 
 computed in the frequency domain.


For SE,
 a beamformer steered to $\hat\theta$ is given by
\begin{equation}
\mathbf{w}_f=\frac{{\mathbf{a}}_f(\hat\theta)}{\|{\mathbf{a}}_f(\hat\theta)\|_2},
\qquad
\hat{s}_{ft}=\mathbf{w}_f^{\mathrm H}\mathbf{x}_{ft},
\label{eq:beamforming}
\end{equation}
where $\mba_f$ can be either the algebraic or measured SV pointing at $\hat{\theta}$ direction. Assuming spatially white noise, choosing $\mathbf{w}
_f\propto\hat{\mathbf{a}}_f$ yields the MaxSINR beamformer~\cite{kim1989signal}.


\vspace{-2mm}
\subsection{Latent diffusion models for image out-painting}
\label{subsec:ldm-theor}
\vspace{-1mm}

Given a cropped image $\mbx^L \in\mathbb{R}^{I_L\times J_L}$, 
 \textit{out-painting} aims 
 at recovering the missing regions 
 of an original image $\mbx^H\in\mathbb{R}^{I_H \times J_H}$ 
 with $I_H > I_L, J_H>J_L$. 
This approach naturally extends 
 for audio super-resolution~\cite{liu2023audiosrversatileaudiosuperresolution}.
State-of-the-art performances on these tasks 
 are obtained with latent diffusion models~\cite{xiao2020image, rombach2022highresolutionimagesynthesislatent,
 ho2020denoisingdiffusionprobabilisticmodels,
 evans2024fasttimingconditionedlatentaudio},
 where the diffusion is processed in a latent space learned 
 by a variational autoencoder (VAE).

Let $E_\phi$ and $D_\psi$ be the VAE encoder and decoder, i.e. $\mathbf{z}_0 = E_\phi(\mathbf{x}^{H}), \hat{\mathbf{x}}^{H} = D_\psi(\mathbf{z}_0)$.
In diffusion models, the forward process represents the variational posterior $q(\mathbf{z}_{1:T}|\mathbf{z}_0)$ as a Gaussian Markov chain, where noise is gradually added to the latent vector $\mathbf{z}_0$ according to a fixed schedule. In latent diffusion, the reverse process $p_\theta(\mathbf{z}_{0:T}|\mathbf{c})$ is similarly modeled as a Gaussian Markov chain, conditioned on a context embedding $\mathbf{c}=\mathcal{C}(\mathbf{x}^L)$ injected via concatenation or cross-attention into the denoising network $\epsilon_\theta(\mathbf{z}_t,t,\mathbf{c})$.
The function $\epsilon_\theta$ is used to predict the mean of each $p_\theta(\mathbf{z}_{t-1}\!\mid\!\mathbf{z}_t,\mathbf{c})$. The training score is $\mathbb{E}\left[ \|\epsilon - \epsilon_\theta(\mathbf{z}_t, t, \mathbf{c}) \right \|^2]$ where $\epsilon \sim \mathcal{N}(\mathbf{0}, \mathbf{I})$ and $\bold{z}_t$ is a linear combination of $\bold{z}_0$ and $\epsilon$.


\vspace{-2mm}
\section{Proposed Method}
\label{sec:method}
\vspace{-1mm}

This section introduces the proposed SV upmixing method 
 that extends a latent diffusion models to SV spatial super-resolution, 
 and discuss its application for audio analysis.

\vspace{-2mm}
\subsection{Steering vector upmixing}
\label{subsec:upmixing}
\vspace{-1mm}

The proposed \texttt{SIRUP} model is a conditioned latent diffusion model that uses the first $M$ channels to generate the missing $M' - M$ ambisonic channels, yielding the upmixed SVs $\hat{\mbA}^\text{up} \in \mathbb{C}^{F \times M'}$ with $M' > M$. Let $\hat{\mathbf A}\!\in\!\mathbb C^{F\times M}$ be the concatenation of
measured FOA SVs across frequency and define the conditioning tensor
by zero-padding channels as
\begin{equation}
\mathbf c=\big[\hat{\mathbf A},\,\mathbf 0_{F\times(M'-M)}\big]
\in\mathbb C^{F\times M'}.
\label{eq:cond}
\end{equation}
Optionally, the algebraic SVs may fill \(\mathbf c_{:,M+1:M'}\) if steering directions are known.

Then we encode the condition with a VAE encoder $E_\phi$ and run latent
diffusion conditioned on this embedding.
Starting from pure noise
$\mathbf z_T\!\sim\!\mathcal N(\mathbf 0,\mathbf I)$, the denoiser
$\epsilon_\theta(\mathbf z_t,t,E_\phi(\mathbf c))$ iteratively produces
$\mathbf z_{t-1}$ until $\mathbf z_0$. The decoder $D_\psi$ then produces
the super-resolved ambisonic SVs:
\begin{equation}
\mathbf z_{t-1}=\mathrm{\epsilon}_\theta(\mathbf z_t,t;E_\phi(\mathbf c)),
\quad
\hat{\mathbf A}^{\text{up}}=D_\psi(\mathbf z_0).
\label{eq:ldm}
\end{equation}

During inference, 
the latent diffusion model 
is used to sample from pure noise and the condition, 
to generate a suitable latent candidate for the decoder. 
The VAE thus acts as a training-time regularizer 
establishing a well-behaved latent manifold, 
through a small KL-divergence penalty. 

To enhance spatial reconstruction, 
 we employ a composite loss function 
 that combines a cosine similarity term 
 with both feature-matching and mean squared error (MSE) objectives.
This significantly 
  improves both learning stability and overall performance.
Furthermore, dilated convolutions are introduced 
  along the frequency axis of the network architecture. 
This modification enforces 
 spatial coherence across different frequency bands, 
 while the microphone axis coherence is already implicitly handled 
 by the cosine similarity term.
 
\subsection{Downstream tasks with up-mixed steering vectors}

Figure~\ref{fig:pipeline} outlines the proposed upmixing pipeline for sound-scene analysis and synthesis. First, measured SVs $\hat{\mba}_f\!\in\!\mathbb{C}^M$ for the target source are estimated from the multichannel mixture $\mbx_{ft}\!\in\!\mathbb{C}^M$ over a short time window using Eq.~\eqref{eq:measured_svect}. Next, the $M$-channel ambisonic representation of $\hat{\mba}_f$ is upmixed to a higher ambisonic order with $M'$ channels via \texttt{SIRUP} (cf.\ Section~\ref{subsec:upmixing}), and then mapped back to the signal domain as SVs of a virtual $M'$-channel spherical array.

SSL is subsequently performed in the upmixed domain
using Eq.~\eqref{eq:ssl} as a grid search using algebraic SVs projected onto the upmixed counterparts. For enhancement and rendering, the upmixed SVs can be truncated to the original $M$-channel subspace to obtain denoised spatial images. Note that \texttt{SIRUP} can also be trained with a denoising objective to refine the first $M$ ambisonic coefficients, thereby improving the accuracy of the estimated SVs.

\section{Evaluation}

\begin{table*}
    \centering
    \footnotesize
    \caption{Performances for different spatial representations, averaged over 30 simulated rooms. Ground-truth HOA values (gray) are provided for reference. Beamwidth is measured at -3dB (lower is better).}
    \begin{tabular}{l S S S S S S S S S}
        \toprule
        & \multicolumn{3}{c}{DI [dB] $\uparrow$} 
        & \multicolumn{3}{c}{3-dB BW [\si{\degree}] $\downarrow$} 
        & \multicolumn{3}{c}{SL [dB] $\downarrow$} \\
        \cmidrule(lr){2-4} \cmidrule(lr){5-7} \cmidrule(lr){8-10}
        & {FOA} & {Upmixed} & {\hoa{HOA}} 
        & {FOA} & {Upmixed} & {\hoa{HOA}} 
        & {FOA} & {Upmixed} & {\hoa{HOA}} \\
        \midrule
        $\mathcal{D}_{\text{RT60}}$  
            & \num{10 \pm 2.6} & {\bfseries \num{19.8 \pm 2.3}} & \hoa{\num{20 \pm 2.2}} 
            & \num{30 \pm 6}  & {\bfseries \num{24 \pm 3.3}}  & \hoa{\num{24 \pm 2}} 
            & \num{-0.9 \pm 0.7} & {\bfseries \num{-9.5 \pm 3.1}} & \hoa{\num{-11.2 \pm 2.8}} \\
        $\mathcal{D}_{\text{SNR}}$ 
            & \num{8.1 \pm 2.7} & {\bfseries \num{17.1 \pm 2.1}} & \hoa{\num{17.7 \pm 2.0}} 
            & \num{48 \pm 6.7} & {\bfseries \num{27 \pm 3.5}} & \hoa{\num{26 \pm 2.2}} 
            & \num{-1.2 \pm 0.9} & {\bfseries \num{-9.6 \pm 3.4}} & \hoa{\num{-11.7 \pm 2.7}}\\
        \bottomrule
    \end{tabular}
    \label{tab:di}    
\end{table*}

\begin{table}[]
    \centering
    \vspace{-1mm}
    \caption{Source synthesis enhancement 
    from mixture across varying for mixtures of two speech sources.}
    \label{tab:beamforming_results}
    \resizebox{\linewidth}{!}{
    \begin{tabular}{l|cc|cc}
    \toprule
    & \multicolumn{2}{c}{Measured SVs} & \multicolumn{2}{c}{Algebraic SVs following SSL}\\
    Metric & SV-FOA & SV-SIRUP-M & SV-alg FOA & SV-alg SIRUP \\
    \midrule
        SDR [dB] & 17.2 ± 3.2 & \textbf{17.4 ± 3.1} & 12.6 ± 7.4 &  \textbf{13.0 ± 7.2}  \\
        SIR [dB] & 38.8 ± 3.6 & 38.8 ± 3.3  & 33.5 ± 7.8 & \textbf{34.0 ± 7.5}  \\
        SAR [dB] & 17.3 ± 3.2 & \textbf{17.4 ± 3.1} & 12.6 ± 7.3 & \textbf{13.0 ± 7.2}  \\
    \bottomrule
    \end{tabular}
    }
\end{table}

This section presents the performance of the proposed upmixing approach 
 in terms of SSL and beamforming metrics. 

\vspace{-1mm}
\subsection{Experimental settings}
\vspace{-1mm}

We first explain the experimental data,
 model configuration, and evaluation metrics.

\vspace{-2mm}
\subsubsection{Experimental data}

Room impulse responses (RIRs) 
 were simulated using the image source model (ISM)
 in the \texttt{pyroomacoustics} library~\cite{Scheibler_2018}.
For each investigation, 
 we generated 30 distinct acoustic scenes 
 with various signal-to-noise ratios ($\mathcal{D}_{\text{SNR}}$) 
 and reverberation times ($\mathcal{D}_{\text{RT60}}$).
The microphone array was centrally 
 placed in a $6 \times 4 \times 3$ \si{\meter} room, 
 with its horizontal position randomized by up to 1 \si{\meter}. 
Source locations were randomly selected 
 within the azimuth plane, with a fixed elevation of approximately 28\si{\degree}. 
This specific elevation was chosen
 to simulate a typical desktop condition, 
 where the microphones are placed on top of a table.
All sources were modeled as point speech sources, 
 randomly selected and cropped to 4 seconds from the
 \textit{dev-clean} folder of the \textit{LibriSpeech} database~\cite{7178964}.

For SSL experiments and testing the upmixing method,
 we simulated rooms containing a single speech source 
 with reverberation and noise. 
The SVs were directly estimated 
 from the mixture using Eq.~\eqref{eq:measured_svect}.
For SE experiments, 
 we simulated mixtures comprising two speech sources 
 across various acoustic environments. 
To compute their SVs, 
 we introduced a 2-second delay 
 between the start times of the two 4-second sources. 
This configuration allowed us to use the first 2 seconds 
 and the last 2 seconds of the overall mixture for the SV estimation 
 of the first and second sources, respectively.

All data were sampled at $f_s = 16$ \si{\kilo\hertz}. 
The short-time Fourier transform (STFT)
 was computed using a 512-sample frame size with 50\% overlap 
 and a Hamming window. 
For the ambisonic representation, 
 third-order data ($M'=16$) were designated as the target HOA data, 
 corresponding to first-order ambisonics ($M=4$) as the low-resolution input.
 
\vspace{-3mm}
\subsubsection{Model configuration}

The training of $\texttt{SIRUP}$ 
 was conducted in a two-stage process.
In the first stage, a
 VAE was trained to reconstruct HOA SVs conditioned on FOA SVs. 
The training data consisted of measured SVs obtained 
 from single-source noisy mixtures 
 utilizing Eq.~\eqref{eq:measured_svect}.
The autoencoder (3.1M parameters) was optimized 
 for 40 epochs using the AdamW optimizer 
 with 
 a learning rate of $3 \times 10^{-4}$. 
The objective function (Section~\ref{sec:method})
 combined an $\ell_2$ reconstruction loss, a cosine loss objective, 
 a perceptual loss, and a small KL term.

The second stage involved freezing the encoder and fine-tuning 
 only the decoder for an additional 20 epochs. 
This fine-tuning utilized a combination of MSE
 and cosine similarity losses, 
 applied with an exponential learning-rate schedule.
We trained an UNet-based latent diffusion model (4.1M parameters)
 as in \cite{evans2024fasttimingconditionedlatentaudio}.
This model operates within the VAE latent space 
 to perform the upmixing task, conditioned on the FOA inputs.
Prior to inputting the latent values to the UNet, 
 the encoder outputs were scaled 
 to the range $[-1, 1]$
 \cite{evans2024fasttimingconditionedlatentaudio}.

 Conditions were injected into the UNet in two ways:
  (i) concatenation of the FOA tensor 
  with the noisy latent representation at the UNet input layer
  and (ii) application of cross-attention 
  within each architectural block.
 Training was conducted for 100 epochs 
  with AdamW at a learning rate of $3 \times 10^{-4}$. 
The total number of diffusion steps, $T$, 
 was set to $1,000$ during training and $200$ during inference.
 
For training,
 3,000 pairs of measured FOA and HOA SVs
 $(\hat{\mbA} \in \mathbb{C}^{F\times M}, \; \hat{\mba}^{\text{up}} \in \mathbb{C}^{F \times M'})$
 were used.
The target HOA data
 were generated by convolving the source signals with the simulated RIRs. 
The corresponding FOA data were
 obtained by retaining only the first $M=4$ channels 
 of the HOA data. 
We represent complex SVs as stacked real/imag parts, hence the leading
dimension “2” in tensors of shape $(2, F, M)$ and $(2, F, M')$, respectively. 
The remaining $(M' - M)$ channels required for the model
 input were addressed by zero-padding the FOA conditioning tensor.
 
\vspace{-2.5mm}
\subsubsection{Evaluation metrics}
\vspace{-1mm}

To assess performance, 
 we utilized two distinct evaluation sets: 
 $\mathcal{D}_{\text{SNR}}$ and $\mathcal{D}_{\text{RT60}}$.
In the $\mathcal{D}_{\text{SNR}}$ set, 
 the SNR was varied within the range $[5, 20]$\,dB
 while the reverberation time was fixed at $\text{RT60}=0.2$\,s. 
Conversely, in the $\mathcal{D}_{\text{RT60}}$ set,
 $\text{RT60}$ was varied within $[0.2, 0.7]$\,s 
 with a fixed $\text{SNR}=20$\,dB.

SSL performance was primarily evaluated in terms of the angular error
 on the azimuthal plane, benchmarked against an SRP baseline derived from FOA mixtures.
To assess the quality of the upsampled SVs, 
 we considered the directivity index (DI),
 the 3dB-beamwidth (3dB BW), and sidelobe (SL) metrics.
Finally, the output of the beamforming process 
 was evaluated using common source separation metrics,
 as detailed in Tab.~\ref{tab:beamforming_results} and computed using the \texttt{bss\_eval} toolkit~\cite{book_vincent}.

\vspace{-1mm}
\subsection{Experimental results}

\begin{figure}[t]
    \centering
    \includegraphics[width=.99\linewidth]{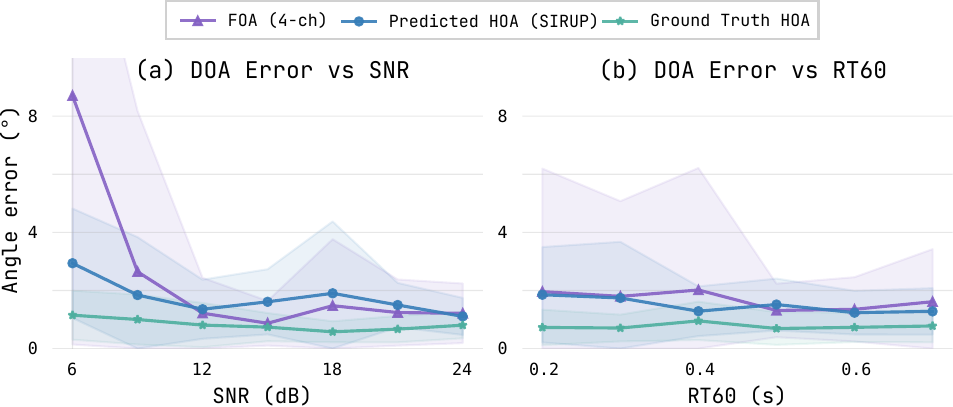}
    \caption{Average angular errors across conditions using different localization methods and SV models, using SRP (Eq. (\ref{eq:beamforming})). 
    }
    \label{fig:angularerror}
\end{figure}

\begin{figure}[t]
    \centering
    \includegraphics[width=.99\linewidth]{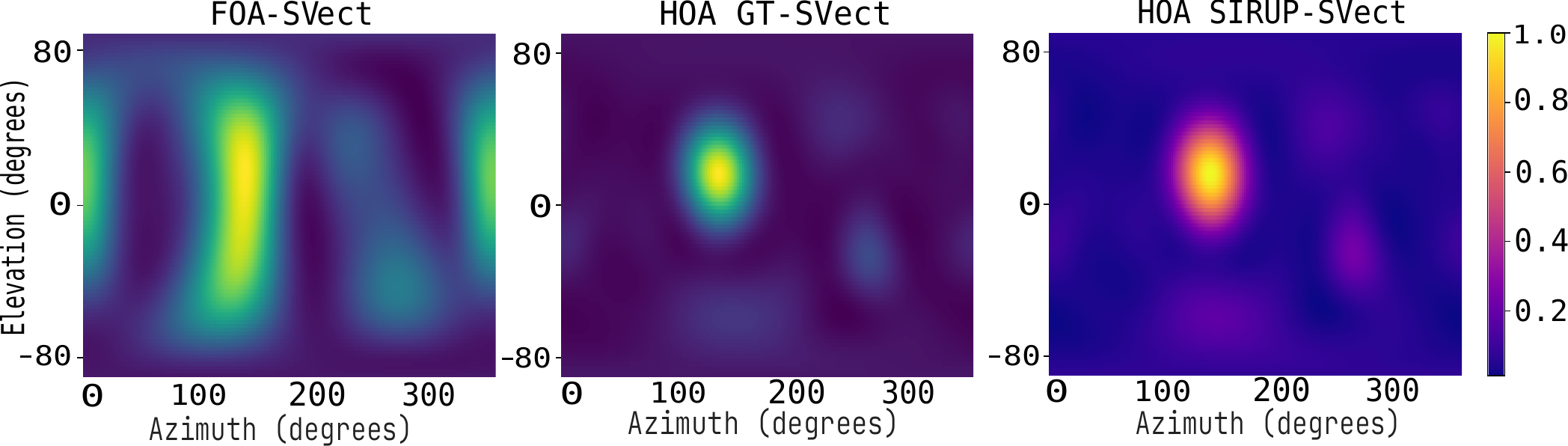}
    \caption{2D Heatmap comparaison of estimated SV for FOA and ground truth HOA setup.}
    \label{fig:heatmap_3d}
    \vspace{-4mm}
\end{figure}

Fig.~\ref{fig:angularerror} reports the average DOA estimation error of the sound source by pickpeaking on the SRP angular map (shown in Fig~\ref{fig:heatmap_3d}, against different noise and reverberation ratios. \texttt{SIRUP} was shown to have outperformed the FOA in noisy conditions, begin closed to the topline using HOA setup. We however reports comparable results increasing the reverberation.

Next, we investigate the spatial quality of the upmixed SVs compared to the FOA baseline in terms of beampattern metrics computed azimuthal plane. As reported in Table \ref{tab:di}, the proposed method \texttt{SIRUP} yields an average beamwidth improvement of $+10$\si{\degree} and sidelobes suppression of $-9$ dB. Qualitatively, this performances translate in narrower beam shapes comparable to the one of a HOA setup, as shown in~\ref{fig:heatmap_3d}.

Finally, we studied the beamforming performance for mixtures of 2 sources. The results reported in Table \ref{tab:beamforming_results} compare the denoised output when using the following SVs model for beamforming as in Eq.~\eqref{eq:beamforming}: the SVs compute for FOA mixture (SV-FOA), the first $M$ channels of the \texttt{SIRUP}'s output (SV-SIRUP-M), the algebraic SVs from DOA estimation of both the measured FOA SVs (SV-alg FOA) and \texttt{SIRUP} (SV-alg SIRUP), respectively. The performances of SV-SIRUP-M are expected to be comparable the one of SV-FOA denoting effective training of the diffusion model. However, we observed improvements, which can be explained by the capacity of the model to denoise SVs during inference. In addition, as we have better DOA estimation using \texttt{SIRUP}, we also note improvements for the SV-alg experiments.
In most practical cases, \texttt{SIRUP} is outperforming the FOA conditions with equal information, leading to spatial-resolution improvements in every experiences.

\vspace{-1mm}
\section{Conclusion}

This paper presented 
a novel diffusion-based approach named \texttt{SIRUP},
 designed to significantly enhance the spatial information 
 derived from FOA setups, 
 allowing it to emulate the performance of HOA microphone arrays. 
\texttt{SIRUP} directly addresses the inherent limitations of FOA systems, 
 such as lower spatial resolution 
 and imprecise SSL, by virtually upmixing SVs 
 using a two-step latent diffusion model.
The experimental results demonstrated that
 \texttt{SIRUP} achieved significant improvements 
 in steering vector upmixing, sound source localization, 
 and speech denoising compared to FOA systems. 

Future work will focus on comprehensively 
 utilizing the upmixed SVs for downstream tasks 
 like source separation and rendering. 
We also plan to make the model capable
 of learning from noisier SVs.
These advancements could significantly enhance
 machine listening applications
 requiring high spatial accuracy, such such as augmented reality, robotics, radar, and autonomous driving systems.
 





\clearpage
{
\small
\bibliographystyle{IEEEbib_abrv}
\bibliography{refs}
}

\end{document}